\documentstyle[12pt]{article}
\begin{document}
%\begin{flushright}
%hep-th/9707004
%\end{flushright}
\vskip .7cm
\begin{center}
{\bf { Quantum Groups, $q$-Dynamics and Raja{\it ji}}}

\vskip 1cm

 {\bf R. P. Malik}
\footnote{ E-mail: malik@boson.bose.res.in  }
\footnote{ Talk given in the ``Raja{\it ji}  Symposium'' (22nd Feb 2001)
held at The Institute of Mathematical Sciences, Madras (India) to honour 
Prof. G. Rajasekaran on the occasion of his 65th birth anniversary.}\\
{\it S. N. Bose National Centre for Basic Sciences,} \\
{\it Block-JD, Sector-III, Salt Lake, Calcutta-700 098, India} \\

\vskip 1cm

\end{center}
{\bf Abstract:}               
We sketch briefly the essentials of the quantum groups and their
application to the dynamics of a $q$-deformed simple harmonic oscillator
moving on a quantum line, defined in the $q$-deformed cotangent
(momentum phase) space. In this
endeavour, the quantum group $GL_{qp} (2)$- and the conventional rotational
invariances are respected together. During the course of this discussion, 
we touch upon Raja{\it ji}'s personality as a critical physicist and a bold and 
adventurous man of mathematical physics.\\
\baselineskip=16pt
---------------------------------------------
\vskip .3cm

The basic idea behind the concept of
``deformation'' in theoretical physics is quite old one.
In fact, the two most successful and well-tested theories of 20th century,
namely; the quantum mechanics and the special theory of relativity, can be 
thought of as the ``deformed'' versions of their ``undeformed'' 
counterparts: the
classical mechanics and the Galilean relativity. The deformation parameters
in these theories are supposed to be the Planck constant ($\hbar$) and the 
speed of light ($c$) [1,2] respectively (which turn out to be the
two fundamental constants of nature). In the limit when $\hbar \rightarrow 0$ 
and $c \rightarrow \infty$, we get back the corresponding ``undeformed'' 
physical theories.  Long time ago, it was proposed
that space-time might become noncommutative [3-5] if we probe the deeper
structure of matter with energies much higher than the typical scale of energy
for quantum mechanics. Nearly a couple of decades ago, this idea got a shot in 
its arms in the context of inverse scattering method (and Yang-Baxter 
equations) applied to the integrable systems [6] and it was conjectured that 
the deformation of groups based on the quasi-triangular Hopf algebras [7] 
together with the ideas 
of noncommutative geometry [8] might provide a ``fundamental length'' 
($l_{p}$) in the context of space-time quantization. This will complete the 
{\it trio} (i.e. $\hbar, c, l_{p}$) of fundamental constants of nature and 
will, thereby, enable us to express physical quantities in terms of these 
natural units. Recently, there has been an upsurge of
interest in the noncommutative spaces [8, 9] in the context of branes in
string theory and matrix model of M-theory. However, we shall discuss here
some aspects of noncommutativity associated with the space-time structure in
the framework of quantum groups {\it alone} and will not touch upon the 
noncommutativity associated with the string/M-theory.

Let us begin with a $q$-deformation
(where $q$ is a dimensionless quantity), introduced as a noncommutativity
parameter for the spacetime coordinates in the $D$- dimensional flat 
Minkowski (configuration) manifold, as 
$$
\begin{array}{lcl}
x_{\mu} x_{\nu} = q \;x_{\nu} x_{\mu}, \;\;\quad\;\;
(\mu, \nu = 0, 1, 2,.....D-1).
\end{array}\eqno (1)
$$
It can be checked that (1) is 
{\it invariant under the Lorentz boost transformations}
iff $ \mu < \nu $. Moreover, if we reduce (1) to a two
dimensional ``quantum plane'' in space (i.e. $\mu = 1, \nu = 2$)
$$
\begin{array}{lcl}
x \;y = q \;y \;x,
\end{array}\eqno (2)
$$
we see that the conventional rotational invariance 
for a two dimensional ``undeformed'' plane is violated. In some sense, 
the homogeneity and isotropy of space-time becomes questionable because
of the loss of these two conventional invariances. In the
limit $q \rightarrow 1$, the ``quantum plane'' reduces to an ordinary plane
with its rotational symmetry intact.

It was a challenging problem to develop a consistent $q$-deformed dynamics
where conventional invariances were respected. In this context, the Lagrangian
and Hamiltonian formulation of a $q$-deformed dynamics was considered 
in the tangent and cotangent spaces,
defined over 2D $q$-deformed configuration space (corresponding to the
definition (2)) [10]. In this approach, however,
the conventional rotational invariance 
was lost and the status of a one-dimensional physical system was not clear. On 
the positive side of this approach, a rigorous $GL_{qp} (2)$ invariant 
differential calculus was developed and then it was applied to the 
construction of a consistent $q$-dynamics.  In another interesting attempt, 
a $q$-deformation was introduced in the Heisenberg
algebra [11]. As a result, it was impossible to maintain the Hermiticity 
property of the phase variables {\it together}. This led to the introduction 
of a new coordinate variable in the algebra. Consequently, a single point 
particle was forced to move on {\it two trajectories} at a given value of 
the evolution parameter for $q \neq 1$ (which was not found to be a 
physically interesting feature).  
In an altogether different approach, a $q$-deformation was introduced 
in the cotangent (momentum phase) space defined over a
one-dimensional configuration manifold [12].  In this endeavour, 
a ``quantum-line'' was defined in the 2D cotangent manifold as
$$
\begin{array}{lcl}
x(t) \;\pi (t) = q \;\pi (t)\; x(t), \qquad
\end{array} \eqno(3)
$$
where $t$ is a commuting real evolution parameter and $x(t)$ and $\pi (t)$ are
the phase space variables. In relation (3), 
the conventional rotational invariance is maintained because a rotation 
{\it does not mix a coordinate with its momentum}. However,
a rigorous differential calculus was not developed in this approach and
{\it dynamics was discussed by exploiting the on-shell conditions alone}. It 
was also required that the solutions to equations of motion should be such that the quantum-line (3) is satisfied for all values of the evolution parameter.
This way of deformation was generalized to the multi-dimensional systems [13-15]
$$
\begin{array}{lcl}
x_{\mu} x_{\nu} = x_{\nu} x_{\mu}, \quad
\pi_{\mu} \pi_{\nu} = \pi_{\nu} \pi_{\mu}, \quad
x_{\mu} \pi_{\nu} = q \;\pi_{\nu} x_{\mu},
\end{array} \eqno(4)
$$
and the dynamics of (non)relativistic systems was discussed by exploiting
the on-shell conditions only. {\it Prof. G. Rajasekaran 
(popularly known as ``Rajaji'' in the physics community of India)
is blessed with a very critical mind. Not only he is critical about
others' work, he is self-critical too. In fact, he was very much critical about
these relations in Eq. (4) and argued that there must be some quantum
group symmetry behind this choice of relations. His criticism spurred
our interest in this problem a great deal}. As a result, we were able
to find that, under the following transformations for the pair(s) of
phase variable(s): $ (x_{0}, \pi_{0}), (x_{1}, \pi_{1})...........(x_{D-1},
\pi_{D-1})$:
$$
\begin{array}{lcl}
&&x_{\mu} \;\;\;\rightarrow\;\;\; A \;x_{\mu} + B \;\pi_{\mu},\nonumber\\
&&\pi_{\mu} \;\;\;\rightarrow\;\;\; C \;x_{\mu} + D \;\pi_{\mu},
\end{array} \eqno(5)
$$
where $A, B, C, D$ are the elements of a $2 \times 2$ 
matrix belonging to the quantum group
$GL_{qp}(2)$ and obeying the braiding relations in rows and columns (with
$ q, p \in {\cal C}/\{0 \}$) as:
$$
\begin{array}{lcl}
&& A B = p \;B A, \quad A C = q \;C A, \quad B C = (q/p)\; C B, \quad
B D = q \;D B,\nonumber\\
&& C D = p\; D C, \quad A D - D A = ( p - q^{-1})\; B C = (q - p^{-1})\; C B,
\end{array}\eqno (6a)
$$
the relations (4) {\it remain invariant for any arbitrary ordering of $ \mu$
and $\nu$ if the parameters of the group are restricted to obey $pq = 1$}.
In fact, relations (4) respect the conventional
Lorentz invariance as well as the quantum group $GL_{q,q^{-1}} (2)$
$$
\begin{array}{lcl}
&& A B = q^{-1}\; B A, \quad A C = q \;C A, \quad B C = q^2\; C B, \nonumber\\
&&B D = q \;D B, \quad
 C D = q^{-1}\; D C, \quad A D = D A,
\end{array}\eqno (6b)
$$
invariance {\it together} if we assume the commutativity of elements
$A, B, C, D$ of the above quantum group
with the phase variables $x_{\mu}$ and $\pi_{\mu}$. It will be noticed 
that the relationship (6b) has been derived from (6a) for $pq = 1$ (and
$ GL_{q, q^{-1}} (2) \neq GL_{q} (2)$). In fact, relations (4) and symmetry
transformations (6a,6b) were exploited for the discussion of a consistent
$q$-dynamics for some physical systems in the multi-dimensional 
phase space [16].

To develop a consistent $q$-dynamics in the 2D phase space for a one 
dimensional simple harmonic
oscillator (1D-SHO), we shall exploit the definition of a quantum-line (3) 
in the phase space. This relationship
remains invariant under the conventional rotations as well as the following
quantum group symmetry transformations
$$
\begin{array}{lcl}
\left (
\begin{array}{c}
x \\
\pi \\
\end{array} \right ) \;\;\rightarrow\;\;
\left ( \begin{array}{cc}
A & B\\
C & D\\
\end{array} \right )
\left (
\begin{array}{c}
x \\
\pi \\
\end{array} \right ), 
\end{array}\eqno (7)
$$
where $A, B, C, D$ are the elements of the quantum group $GL_{qp}(2)$ that obey
relations (6a). It will be noticed that the quantum-line (3) is also invariant
under transformations corresponding to the quantum group $GL_{q}(2)$ which
is endowed with elements ($ A, B, C, D)$ 
{\it that obey relations (6a) for $q = p$} [13]. 
In fact, both these quantum groups possess identity, inverse, closure property 
and associativity under the binary operation 
$(\cdot)$ as the {\it matrix multiplication}.
However, these quantum groups form what are known as 
{\it pseudo-groups}. To elaborate this point,
let us examine the group properties of the simpler group $GL_{q}(2)$. The
identity element ($I$) of this group can be defined from its 
typical general element ($T$) as:
$$
\begin{array}{lcl}
T_{ij} =
\left ( \begin{array}{cc}
A & B\\
C & D\\
\end{array} \right ) \in GL_{q} (2),
\quad \mbox{and} \quad \; T_{ij} \rightarrow \delta_{ij} \equiv I =
\left ( \begin{array}{cc}
1 & 0\\
0 & 1\\
\end{array} \right ), 
\end{array}\eqno (8)
$$
for $ A = D = 1$ and $ B = C = 0$. The determinant of $T$ 
($ det\; T$) turns out to be the
central for this group in the sense that it commutes with all the elements
($ A, B, C, D$)
$$
\begin{array}{lcl}
&& det\; T = A D - q B C = D A - q^{-1} C B, \nonumber\\
&& (det \;T)\; ( A, B, C, D ) = ( A, B, C, D )\; (det\; T),
\end{array} \eqno(9)
$$
as can be seen from the $q$-commutation relations of elements $A, B, C, D$ 
belonging to the quantum group $ GL_{q} (2)$. Now the inverse of $T$ can be 
defined as
$$
\begin{array}{lcl}
T^{-1} = \frac{1} { A D - q B C}\;
\left ( \begin{array}{cc}
D & - q^{-1} B\\
- q C & A\\
\end{array} \right ) \equiv \frac{1}{ D A - q^{-1} C B}\;
\left ( \begin{array}{cc}
D & - q^{-1} B\\
- q C & A\\
\end{array} \right ), 
\end{array}\eqno(10)
$$
because $ T \cdot T^{-1} = T^{-1} \cdot T = I$. The closure property 
($ T \cdot T^{\prime} = T^{\prime\prime}$), under matrix multiplication,
can be seen by taking two matrices $T$ and $T^\prime \in GL_{q}(2)$ and
demonstrating 
$$
\begin{array}{lcl}
\left ( \begin{array}{cc}
A & B\\
C & D\\
\end{array} \right ) \cdot
\left ( \begin{array}{cc}
A^\prime & B^\prime\\
C^\prime & D^\prime\\
\end{array} \right ) =
\left ( \begin{array}{cc}
A^{\prime\prime} & B^{\prime\prime}\\
C^{\prime\prime} & D^{\prime\prime}\\
\end{array} \right ) \in GL_{q}(2),
\end{array}\eqno(11)
$$
where it is assumed that elements of the above two matrices, belonging to the
quantum group $GL_{q}(2)$, commute among themselves (i.e.,
$ [ T_{ij}, T_{kl}^{\prime} ] = 0$). Exploiting the closure property,
it can be checked that the associativity property is also satisfied, i.e.,
$$
\begin{array}{lcl}
T \cdot \bigl ( T^\prime \cdot T^{\prime\prime} \bigr ) =
 \bigl ( T \cdot T^\prime \bigr ) \cdot T^{\prime\prime}.
\end{array}\eqno(12)
$$
The requirement that $ [ T_{ij}, T^{\prime}_{kl} ] = 0$, entails upon quantum
group (e.g. $GL_{q}(2)$) to be a {\it pseudo-group}. This is because of
the fact that: (i) the elements of the product $T^2 = T \cdot T$ do not
belong to the quantum group (e.g. $GL_{q}(2)$), and (ii) the condition
$ [ T_{ij}, T_{kl}^{\prime} ] = 0$ is not taken into account while defining
the inverse matrix $(T^{-1})$ where the elements of $T^{-1}$ and $T$ are
taken to be non-commuting in the proof: $ T \cdot T^{-1} = T^{-1} \cdot T = I$.

In our work [16], a consistent $q$-dynamics was developed where conventional
rotational (and/or Lorentz) symmetry
invariance, together with a quantum group symmetry
invariance, was maintained. A $GL_{qp}(2)$ invariant differential calculus, 
consistent with the Yang-Baxter equations, was developed in 2D 
(momentum) phase space and then it was applied for the discussion
of dynamics of some physical systems. As an example here,
we begin with the following Lagrangian for the 1D-SHO [16]
$$
\begin{array}{lcl}
L = P \;\dot x^2 - Q \;x^2,
\end{array}\eqno (13)
$$
where $\dot x, x $ are the velocity and position variables (with $ x \dot x
= pq \;\dot x x$) and $P$ and $Q$ are the parameters which are, 
in general, non-commutative.
A general discussion for the least action principle leads to the 
definition and derivation of the canonical momentum ($\pi$)
and the Euler-Lagrange equation of motion (EOM) as
$$
\begin{array}{lcl}
\pi = \frac{1}{p}\; \frac{\partial L}{\partial \dot x}, \quad
\dot \pi = \frac{1}{p}\; \frac{\partial L}{\partial  x}.
\end{array}\eqno(14)
$$
The differential calculus (with $ dx = \dot x\; dt, d \pi = \dot \pi \;dt$
where $t$ is a real commuting evolution parameter) leads to the derivation
of the following basic relations [16]
$$
\begin{array}{lcl}
&& x \;\pi = q \;\pi \;x, \quad \dot x \;\dot \pi = q \;\dot \pi \;\dot x, 
\quad \pi\; \dot x = p \;\dot x\; \pi, \nonumber\\
&& \pi \;\dot \pi = p q \;\dot \pi\; \pi, \qquad
x \;\dot \pi = q \;\dot \pi \;x + (pq - 1)\; \dot x \;\pi,
\end{array}\eqno(15)
$$ 
which restrict any arbitrary general Lagrangian ($L (x, \dot x)$) to satisfy 
$$
\begin{array}{lcl}
&& x \;\frac{\partial L}{\partial \dot x} = q \;
 \frac{\partial L}{\partial \dot x} \;x, \quad 
\dot x \;\frac{\partial L}{\partial x} = q 
 \;\frac{\partial L}{\partial x}\; \dot x, \quad 
\frac{\partial L}{\partial \dot x}\; \dot x = p\;
 \frac{\partial L}{\partial \dot x} \;\dot x, 
\nonumber\\
&&  \frac{\partial L}{\partial \dot x}  
  \;\frac{\partial L}{\partial  x} = p q\; 
  \frac{\partial L}{\partial  x}  \;
 \frac{\partial L}{\partial \dot x}, \quad 
x \;\frac{\partial L}{\partial x} = q \;
 \frac{\partial L}{\partial x} \;x + (pq - 1)\; \dot x 
 \;\frac{\partial L}{\partial \dot x}.
\end{array}\eqno(16)
$$

We demand that the Lagrangian (13) should satisfy all the basic conditions
listed in (16). As it turns out, there are two interesting sectors of dynamics
for 1D-SHO, described by (13). These are: (i) when the parameters of 
deformations are restricted to satisfy $ pq = 1$, and (ii) when $ pq \neq 1$
for the discussion of solutions to EOM. For the former
case, consistent with the differential calculus [16], we obtain the following
$q$-commutation relations
$$
\begin{array}{lcl}
x \;\dot x = \dot x\; x, \quad P\; Q = Q\; P, \quad \xi\; P = q \;P\; \xi, 
\quad \xi\; Q = q \;Q\; \xi,
\end{array}\eqno(17)
$$ 
where $\xi$ stands for $ x , \dot x$ (i.e. $ \xi = x, \dot x)$. 
Exploiting these relations, we obtain the following EOM for the system under
consideration
$$
\begin{array}{lcl}
\ddot x = - P^{-1} Q\; x \equiv - \omega^2\; x, \qquad (\omega^2 = P^{-1} Q),
\end{array}\eqno(18)
$$ 
which has its solution, at any arbitrary value of the 
evolution parameter $t$, as
$$
\begin{array}{lcl}
x(t) = e^{i \omega t}\; A + e^{- i \omega t}\; B,
\end{array}\eqno(19)
$$
where $A$ and $B$ are the non-commuting constants which can be fixed in terms of
the initial conditions of the dynamics, as given below
$$
\begin{array}{lcl}
A = \frac{1}{2} \;\bigl [\; x (0) + \omega^{-1} \dot x (0)\; \bigr ], \quad
B = \frac{1}{2} \;\bigl [ \;x (0) - \omega^{-1} \dot x (0) \;\bigr ].
\end{array}\eqno(20)
$$ 
The consistency requirements of $GL_{qp} (2)$ invariant differential 
calculus [16] vis-a-vis relations (17) and (20), lead to
$$
\begin{array}{lcl}
A B = B A, \quad B \omega = \omega B, \quad A \omega = \omega A,
\quad x \omega = \omega x, \quad \dot x \omega = \omega \dot x.
\end{array}\eqno(21)
$$
Furthermore, it can be checked that all the $q$-commutation relations [16]
among $x, \pi, \dot x, \dot \pi$ are satisfied at any arbitrary value of $t$.
{\it Thus, we have a completely consistent dynamics in a noncommutative 
phase space for $ pq = 1$.} We have paid a price, however. 
As it has turned out, all the
variables (i.e. $A, B, \omega, x(0), \dot x(0)$),  present in the solution
$x(t)$, are commutative in nature {\it like we have in conventional classical
mechanics}. Thus, in some sense, the nature of this 
dynamics is {\it trivial}. In
other words, as far as the evolution of the system is concerned, we do not see
the effect of $q$-deformation (or noncommutativity) on the dynamics.

This is the point where {\it boldness and adventure of ``Rajaji'' (as a 
mathematical physicist) came to the fore. He argued that we must find out a 
non-trivial solution to the equation of motion where parameters of the
deformation are not restricted to unity (i.e, $ pq \neq 1$)}. Thus, let us 
consider this non-trivial sector of dynamics for 1D-SHO. The Lagrangian (13)
{\it satisfies all but one relations in (16) }
with the following $q$-commutation relations
$$
\begin{array}{lcl}
P Q = Q P, \quad \xi\; Q = p q^2 \;Q \;\xi, \quad  P\; \xi = p \;\xi\; P,
\quad (\xi = x, \dot x).
\end{array}\eqno(22)
$$
The problematic relation of (16) (the last one!) forces us to require:
$$
\begin{array}{lcl}
(pq - 1) \bigl ( Q x^2 - p q\; P \dot x^2 \bigr ) = 0,
\end{array}\eqno(23)
$$
which emerges from the basic condition: $ x \dot \pi = q \;\dot \pi x
+ (pq - 1) \;\dot x \pi$ of Eq. (15). The EOM for the system in this sector
(where $ pq \neq 1$)
$$
\begin{array}{lcl}
\ddot x = - \frac{1} {p^{2} q^{2}}\; P^{-1} Q \;x \equiv - \omega^2 \; x,
\quad (\omega^2 = \frac{1}{p^2 q^2}\; P^{-1} Q),
\end{array}\eqno(24)
$$
has the following general solution in terms of constants $A, B$ and $\omega$
$$
\begin{array}{lcl}
x(t) = e^{i\omega t} \;A + e^{-i\omega t}\; B.
\end{array}\eqno(25)
$$
However, the {\it restriction (23) is satisfied if and only if:  either
$A = 0$ or $ B = 0$}. With the expression for $A$ and $B$, given in (20),
this leads to the restriction:  $ \dot x (0) = \pm \;\omega \;x(0)$. 
Now the general form of the {\it exponential} evolution for the
1D-SHO is either
$$
\begin{array}{lcl}
x(t) = e^{i\omega t}\; A, \quad \mbox {with} \quad A \omega = p q \;\omega A,
\end{array}\eqno(26)
$$
or,
$$
\begin{array}{lcl}
x(t) = e^{-i\omega t}\; B, \quad \mbox {with} \quad B \omega = p q \;\omega B.
\end{array}\eqno(27)
$$
As a consequence, the dynamics for $ pq \neq 1$ does not evolve in the 2D
(velocity) phase space but degenerates into a restricted 1D region. 
In this region, both the restrictions: $ \dot x(t) = \pm \;\omega \;x(t)$ and 
$ Q \;x^2 = p q \;P \;\dot x^2$ are satisfied for all values of the
evolution parameter $t$. In fact, it can
be checked that the evolution of the system, described by Eq. (26) and/or 
Eq.(27), is such that these conditions are very precisely satisfied.

It will be noticed that, unlike the dynamical sector for $ pq = 1$,
here the constants $\omega$ and $A$ (or $B$) do not commute with each-other
and still there exists a consistent ``time'' evolution
for the system, {\it albeit a restricted one}.
In fact, as  a result of the $q$-deformation, the evolution of the
1D-SHO is {\it strictly} on a  1D ``quantum-line'' 
{\it even-though the whole (velocity) phase space} is allowed for
its evolution
\footnote{ The discussion of dynamics in the cotangent (momentum phase) space
has been carried out in the Hamiltonian formulation as well [16].}.
{\it This new feature is completely different from the discussion of a 1D-SHO 
in the framework of classical mechanics}. It will be a nice
idea to discuss the supersymmetric version of this system in the framework of
$q$-deformed dynamics. It will be very interesting to explore the possibility 
of $\hbar$-deformation over $q$-deformation and look for the new aspects
of dynamics when $q$ and $\hbar$  both are present.

\end{document}